# Online Networks, Social Interaction and Segregation: An Evolutionary Approach

Angelo Antoci[1]

Fabio Sabatini[2]

Francesco Sarracino [3]

This version: March 14, 2016

**Abstract**

We have developed an evolutionary game model, where agents can choose between two forms of social participation: *interaction via online social networks* and *interaction by exclusive means of face-to-face encounters*. We illustrate the societal dynamics that the model predicts, in light of the empirical evidence provided by previous literature. We then assess their welfare implications. We show that dynamics, starting from a world in which online social interaction is less gratifying than offline encounters, will lead to the extinction of the sub-population of online networks users, thereby making Facebook and alike disappear in the long run. Furthermore, we show that the higher the propensity for discrimination between the two sub-populations of socially active individuals, the greater the probability that individuals will ultimately segregate themselves, making society fall into a social poverty trap.

[1] Department of Economics and Business, University of Sassari, Italy. Email: antoci@uniss.it. The research of Angelo Antoci has been financed by *Regione Autonoma della Sardegna* (L. R. n. 7, 2007; research project *Capitale sociale e divari economici regionali*).

[2] Department of Economics and Law, Sapienza University of Rome, Italy. Email: fabio.sabatini@uniroma1.it.

[3] Institut national de la statistique et des études économiques du Grand-Duché du Luxembourg (STATEC), Agence pour la normalisation et l'économie de la connaissance (ANEC), and LCSR National Research University Higher School of Economics, Russian Federation. Email: Francesco.Sarracino@statec.etat.lu.

# 1. Introduction

Social interactions affect a variety of behaviors and economic outcomes, including the formation of opinions and tastes, investment in human capital, access to jobs and credit, social mobility, subjective well-being and the emergence of collective action, to name a few. While face-to-face interactions have reportedly been declining in many countries over the last two decades (Putnam, 2000; Costa and Kahn, 2003; Bartolini and Sarracino, 2014; Sarracino and Mikucka, 2016), participation in social networking sites (SNS), such as Facebook and Twitter, has steeply risen (Duggan et al., 2015)[4]. The advent of online social networks has radically changed the way that we interact with others and this change can have major economic and welfare consequences.

In *Bowling Alone*, Putnam (2000) suggested that technology-based private entertainment, such as television, could replace face-to-face meetings and civic engagement in individual preferences. This claim was supported by virtually any empirical test on the role of television, which was found to displace encounters with friends, associational activities and political participation (e.g., Bruni and Stanca, 2008). Following Putnam's argument about television, early Internet studies advanced the "crowding-out hypothesis", according to which, the Internet use crowds-out social engagement. As television, a unidirectional mass medium, displaced so many activities, it stands to reason that the Internet, which allows for interactive communication, might induce a more powerful substitution effect (DiMaggio et al., 2001). The first empirical studies of the relationship between Internet use and face-to-face interactions supported the crowding-out hypothesis (Kraut et al., 1998; Nie et al., 2002). Subsequent studies, on the other hand, found conflicting results, suggesting that the effect of Internet use may vary with users' preferences and personal characteristics (see, for example, Gershuny, 2003; Uslaner, 2004). Yet, these studies are not conclusive: at the time of early studies, using the Internet was predominantly a solitary activity that was connected with private entertainment. The advent of online social networks radically transformed the way that people use the Internet, which largely extended the possibilities to interact with others.

Despite the extent of the transformations brought by online networking, existing research on the relationship between face-to-face interaction and SNS-mediated interaction is limited. There are empirical studies on the effect of broadband access on outcomes like social participation and voting behavior (e.g., Bauernschuster et al., 2014; Falck et al., 2014). A few authors specifically addressed the role of SNS in some aspects of social capital such as face-to-face interaction and trust (Sabatini and Sarracino, 2014a; 2015). These works put the crowding-out hypothesis into perspective, suggesting that face-to-face and Internet-mediated interaction may rather be complementary.

---

[4] Hereafter, online social networks, social networking sites (or SNS) and online networking will be used as synonyms for the sake of brevity. For a discussion about definitions, see Ellison and Boyd (2013).

Additionally, while early sociological studies implicitly suggested the risk of segregating the two populations of Internet users and socially active individuals, more recent works illustrate the emergence of two main types of social actors: those who only interact with others face-to-face and those who develop their social life both online and through face-to-face interactions (e.g., Bauernschuster et al., 2014; Helliwell and Huang, 2013; Sabatini and Sarracino, 2014a). Theoretical studies suggested that the two forms of interaction might be linked to the extent to which communicating via the Internet might allow individuals to cope with distance and busyness in the preservation of their social life (Antoci et al., 2012a; 2014; 2015).

In addition, a third population of socially isolated individuals who devote an increasing share of their time to work and private consumption seems to be growing in richer and emerging countries (Putnam, 2000; Bartolini and Sarracino, 2014; Bartolini and Sarracino, 2015). Antoci et al. (2012b) showed how the choice of social isolation might be a rational response, allowing individuals to adapt to the relational poverty of the surrounding environment and reduction in leisure time.

To date, however, we lack a theoretical framework to study how social interaction via SNS relates to interaction via physical encounters and the intentional withdrawal from social participation, as feared by Putnam (2000) in *Bowling Alone*.

We add to previous literature by developing an evolutionary game model of SNS-mediated social interaction. In our simplified framework, agents can choose between social interaction and social isolation. The latter may be viewed as a drastic form of adaptation to conditions of social decay, increasing busyness and declining opportunities for social engagement, which provides constant payoffs that are independent from the behaviors of others.

Following descriptive evidence concerning Italy and the United States (Sabatini and Sarracino, 2014b; Duggan et al., 2013), we assume that socially active individuals can develop their social relationships face-to-face or through online social networks. Those choosing social interaction can adopt two alternative strategies: 1) to interact both by means of SNS and face-to-face encounters; 2) not to use SNS and to only develop social relationships by means of face-to-face encounters. The distinctive trait of these two strategies is the use of SNS[5].

The analysis shows that, depending on the configuration of payoffs and initial distribution of the three strategies in the population, different Nash equilibria can be reached. In particular, we found that the stationary state, where all individuals choose isolation, is always locally attractive. Thus, it represents a *social poverty trap*, i.e., an equilibrium, where no one has an interest in interacting with others and everybody devotes all of their available time to work or private consumption.

---

[5] We do not use other tools for online communication, such as emails and voice systems (e.g., Skype), for defining the possible strategies of social participation. This is because such tools are commonly spread across the sub-population of socially active individuals, independently of their use of online social networks. Descriptive statistics from various institutions report that virtually the entire population of online adults uses non-SNS-mediated tools of online communication. Distinguishing them from other types of online socially active individuals would make no sense. This aspect will be further explained in Section 2.1.

Only the stationary states in which all individuals play the same strategy can be attractive Nash equilibria. The dynamics leading to those states are self-feeding, to the extent to which agents get a higher payoff when they interact with agents that adopt the same strategy. When the three stationary states are simultaneously attractive, the social poverty trap is always Pareto dominated by the other equilibria and, therefore, it can be considered as the worst-case scenario. However, the possibility of interacting via SNS offers individuals a coping response that allows "defending" their social life from increasing busyness and a reduction in leisure time. This can lead society to an equilibrium in which all agents develop their social relationships through a strategy that encompasses participation in online social networks. Depending on the configuration of parameters, this may be the second best scenario, Pareto dominated by the equilibrium in which everyone interacts by exclusive means of physical encounters. In this case, the widening of the agents' opportunity set for social interactions can prevent the achievement of the first best scenario. At the same time, however, it allows society to avoid the worst-case scenario of the attractive social poverty trap. In all cases, the achievement of a specific equilibrium depends on the initial distribution of the three ways of social interaction in the population.

In this scenario, the propensity for discrimination of socially active individuals defines the structure of the basin of attraction of the social poverty trap[6]. The higher the propensity for discrimination, the greater the probability that individuals will ultimately segregate themselves, making society fall into the trap.

Our contribution is related to at least three literatures. The first literature includes empirical studies that documented a decline in face-to-face social participation in many countries (Putnam, 2000; Costa and Kahn, 2003; Bartolini et al., 2013; Bartolini and Sarracino, 2015). It also includes theoretical studies that explain such a decline in connection with the negative externalities of growth (Bartolini and Bonatti, 2008; Antoci et al., 2012b; 2013).

The second literature is by economists who theoretically and empirically analyzed how Internet use may affect social capital (Campante et al., 2013; Falck et al., 2014; Bauernschuster et al., 2014; Antoci et al., 2012a; 2014; Sabatini and Sarracino, 2014a; 2015). Antoci et al. (2012a; 2014) modeled the choice between two ways of social participation, respectively based on Internet-mediated and face-to-face interaction, in a framework where the time available for social participation is exogenously given. Antoci et al. (2015a) added to previous work by including the choice to withdraw from social participation. The evolutionary framework that is presented in this paper contributes to this body of research in several ways. First, we introduce a new specification of the social interaction mechanism (Section 2.1) that determines the probability of meeting between individuals belonging to each of the three sub-populations that we account for. Second, the resulting configuration of payoffs (Section 2.2) –which allows the outcomes of interaction to vary according to the type of agent with which people are matched–takes into account the propensity for discrimination, allowing us to study its dynamic outcomes in terms of segregation.

---

[6] The classification of dynamic regimes is illustrated in Section 5.

The latter aspect links our work to a third literature that refers to theoretical studies on social interaction and segregation. Schelling's (1969; 1971) seminal contribution explained how people's preference for interaction with similar others –and, therefore, for discrimination of different others– generates dynamics that naturally lead to segregation. Bischi and Merlone (2011) developed Shelling's work by formalizing a two-dimensional dynamic system to study segregation. The authors showed how adaptive rules shape evolutive paths that lead to the emergence of different collective behaviors in the long run. When members of a population are characterized by a limited tolerance of diversity, the complete separation of different populations may occur. Radi et al. (2014a; 2014b) further developed this framework by analyzing the role of regulating institutions constraining the number of individuals of a population that are allowed to enter and exit the system. Our work adds to this literature by illustrating how the configuration of payoffs drives population dynamics towards segregation. If we allow for a configuration of payoffs that reflects a preference for interaction with similar others, then dynamics will lead to the complete separation of the three populations accounted for in our framework. This is consistent with Bischi and Merlone (2011).

The rest of the paper is organized as follows. In section 2, we describe the model and analyze the evolutionary dynamics. Sections 4 and 5 present the basic results and classification of dynamic regimes. Section 6 discusses the possible dynamics predicted by the model for a specific distribution of the different forms of participation suggested by the existing empirical literature. Section 7 concludes.

**2. The Model**

*2.1 The Social Interaction Mechanism*

We consider an economy that is made up of a continuum (of measure one) of identical individuals. In each instant of time *t*, each individual has to choose one of the pure strategies of social interaction that are mentioned in the introduction of this paper:

1) Interaction via online social networks and face-to-face. We call this strategy *SN* (because its distinctive trait is the use of Social Networks). The *SN* strategy entails different degrees of SNS-mediated interaction according to individual preferences. In general, we think of *SN* agents as individuals who develop social ties via SNS at their convenience—for example, by using Facebook to stay in touch with friends and acquaintances or for establishing contacts with unknown others—and meet their contacts in person whenever they want or have time.

2) Interaction by exclusive means of face-to-face encounters. We call this strategy *NS* (because its players make No use of Social networks). The empirical evidence shows that, despite the steep rise in the use of SNS, a remarkable amount of online adults chooses not to use them.

We take for granted that both strategies encompass the use of other non-SNS mediated online tools for social interaction to various extents (e.g., Internet calls, emails, etc.). The distinctive trait of the two strategies is the use of SNS for social interaction, which has two remarkable features: it allows asynchronous and complex interactions; it generates club effects that may favor segregation to the extent to which users get different payoffs when dealing with other users or with non-users.

3) Social isolation, i.e., a strategy in which agents who prefer to devote all of their time to work and to forms of private consumption that do not entail any significant relationship, neither online nor face-to-face (Antoci et al., 2015a). We call this strategy *NP* (for No Participation). *NP* players tend to replace relational goods (e.g., playing a chess tournament with friends) with material goods (e.g., a software for playing chess with a computer). We assume that *NP* agents do not retire from work and that their social relations are limited to on the job interactions.

The withdrawal from social interactions modeled with the *NP* strategy may be viewed as a drastic form of adaptation to conditions of social decay that make *NP* players' payoff constant and completely independent from the behavior of others. The notion of defensive choices is not new in the literature. Hirsch (1976) was the first to introduce the concept of defensive consumption induced by negative externalities of growth. This kind of consumption may occur in response to a change in the physical or social environment: "If the environment deteriorates, for example, through dirtier air or more crowded roads, then a shift in resources to counter these "bads" does not represent a change in consumer tastes but a response, on the basis of existing tastes, to a reduction in net welfare" (Hirsch, 1976, p. 63). Antoci et al. (2012a; 2013) generalized the study of defensive consumption choices to the case of a deteriorating social environment. If the social environment deteriorates, for example, in relation to a shift in prevailing social values or decline in the opportunities of social engagement, then individuals might want to replace the production and consumption of relational goods with the production and consumption of private goods[7]. The authors suggested that the reduction in the time available for social participation could trigger self-feeding processes, leading to the progressive erosion of the stock of social capital.

---

[7] A peculiarity of relational goods is that it is virtually impossible to separate their production from consumption, since they coincide (Gui and Sugden, 2005).

We indicate the shares of individuals playing strategies *SN*, *NS* and *NP* at time *t* with, respectively, $x_1(t), x_2(t), x_3(t)$, where $x_1, x_2, x_3 \geq 0$ and $x_1 + x_2 + x_3 = 1$ hold. The social interaction mechanism, which determines the payoffs of each strategy, is described by the following assumptions:

1) In each instant of time *t*, a share $l \in (0,1)$ of the sub-population of size $x_1$ and of the sub-population of size $x_2$ of individuals respectively playing the *SN* and the *NS* strategy are enjoying their leisure time, which coincides, by assumption, with their social participation time. We say that these individuals are in "*Lmode*". The remaining share of such sub-populations, 1–*l*, is currently working or engaged in private activities that have no (positive or negative) effect on the payoffs of other individuals. We say that these individuals are in "*Wmode*". All of the individuals choosing the *NP* strategy, in the instant of time *t*, are in *Wmode*.

2) In each instant of time *t*, a share $n \in (0,1)$ of the sub-population of size $lx_1$, composed of *Lmode*–individuals playing the *SN* strategy, is interacting online via a social networking site with individuals of the same type, while a share 1-*n* of such sub-population is interacting via face-to-face encounters both with individuals of the same type and individuals belonging to the sub-population of size $lx_2$, composed of *Lmode* -individuals who adopt the *NS* strategy.

3) In each instant of time *t*, an individual choosing either the *SN* or the *NS* strategy has an *l* probability of being an *Lmode*-individual and a 1–*l* probability of being a *Wmode*-individual.

In addition, an *Lmode*-individual adopting the *SN* strategy has a *n* probability of interacting online via a social networking site with individuals of the same type (i.e., *Lmode*-individuals playing *SN* and interacting online via a SNS) and a 1-*n* probability of interacting via face-to-face encounters with individuals of the same type (i.e., *Lmode*-individuals choosing *SN* and interacting via face-to-face encounters) and with *Lmode*-individuals playing the *NS* strategy.

4) In each instant of time *t*, *Lmode*-individuals playing the *NS* strategy interact with *Lmode*-individuals playing the same strategy and with the share 1-*n* of the sub-population of *Lmode*-individuals playing the *SN* strategy and currently interacting via face-to-face encounters.

The values of the shares *l* and *n* are assumed to be exogenously determined (i.e., *l* and *n* are parameters of the model).

5) In each instant of time *t*, every *Wmode*-individual obtains the payoff α, where α is a strictly positive parameter, independently from the strategy (either *SN*, *NS* or *NP*) she adopts, and from the distribution $x_1, x_2, x_3$ of the strategies in the population. The social interaction between *L*mode-individuals and *W*mode-individuals gives players a payoff that is equal to 0.

*2.2 Expected Payoffs*

We assume that the expected payoff of an individual adopting the *SN* strategy is given, in each instant of time *t*, by:

$$EP_{SN}(x_1, x_2) = (1-l)\alpha + ln(\beta ln x_1) + l(1-n)[\gamma l(1-n)x_1 + \delta l x_2] =$$
$$= (1-l)\alpha + \beta l^2 n^2 x_1 + \gamma l^2 (1-n)^2 x_1 + \delta l^2 (1-n) x_2$$

Where:

a) $1-l$ and $l$ are, respectively, the probabilities to be a *Wmode*-individual and an *Lmode*-individual, while $ln$ is the conditional probability to be an *Lmode*-individual (this happens with probability *l*) interacting online via a SNS (this happens with probability *n*). Any individual of this type interacts via SNS with the sub-population of expected size $ln x_1$ of individuals of the same type. The parameter $\beta$ measures the benefit due to this type of interaction. It is important to note that, according to this game framework, the value of *ln* measures either the probability to be an *Lmode*-individual interacting online via SNS or the expected share of *Lmode*-individuals interacting online via SNS in the sub-population (of size $x_1$) of individuals playing *SN*. The parameter $\alpha$ measures the benefit from being in *Wmode* and, therefore, $(1-l)\alpha$ represents the expected benefit deriving from working activity.

b) $l(1-n)$ is the probability to be an *Lmode*-individual playing the *SN* strategy (this happens with probability *l*) and interacting via face-to-face encounters (this happens with probability 1-*n*). Any individual of this type interacts with the sub-population of the expected size $l(1-n)x_1$, of individuals of the same type and with the sub-population of the expected size $l x_2$ of *Lmode*-individuals playing the *NS* strategy; $\gamma$ and $\delta$ are parameters measuring, respectively, the benefits of these two types of interaction.

Analogously, the expected payoff of an individual adopting the *NS* strategy is given, in each instant of time *t*, by:

$$EP_{NS}(x_1, x_2) = (1-l)\alpha + l[\varepsilon l(1-n)x_1 + \eta l x_2] =$$
$$= (1-l)\alpha + \varepsilon l^2 (1-n)x_1 + \eta l^2 x_2$$

Where $\eta$ and $\varepsilon$ are parameters measuring the benefits of an *Lmode*-individual adopting *NS* due to the face-to-face interaction with, respectively, the sub-population of expected size $l x_2$ of individuals of the same type and sub-population of the expected size $l(1-n)x_1$ of *Lmode*-individuals adopting *SN* and interacting via face-to-face encounters.

Finally, the expected payoff of an individual adopting the *NP* strategy is given, in each instant of time *t*, by:

$$EP_{NP} = \alpha > 0$$

The description of payoffs highlights some points about discrimination. First, a clear separation occurs between those who choose to withdraw from social interaction and all the other players. In a sense, *NP* players choose to segregate themselves from the rest of the population. Second, when *SN* players spend their leisure time interacting via SNS, they *de facto* segregate themselves from the two populations of *NS* and *NP* players, who do not use online social networks.

The sub-populations of individuals playing *SN* and *NS* can only meet in the context of face-to-face interactions. The two extreme cases $\delta \lessapprox 0$ and $\varepsilon \lessapprox 0$ entail discrimination. In these cases, in fact, when individuals adopting different strategies of participation meet face-to-face, they get a null or a negative reward. As a result, they will prefer to interact with individuals of their same type. For example, *SN* players may want to check what happens in their online networks while having dinner with friends. *NS* players, who are not familiar with SNS, may, in turn, feel uncomfortable sitting at a table where everyone is checking a smartphone instead of talking to each other. If this is the case, the benefits $\varepsilon$ of the dinner for *NS* players may be null or negative. At the same time, the impossibility to check Facebook during face-to-face interactions –due, for example, to the moral obligation to talk– can make *SN* players feel uncomfortable and anxious (e.g., Shu-Chun et al., 2012). In this case, the benefits $\delta$ of the dinner may be poor or even null or negative for *SN* players too. As a result, *SN* and *NS* players might want to discriminate each other in face-to-face interactions. More generally, players' preference for their similar type maybe interpreted as a matter of homophily. Empirical literature has shown that informal segregation spontaneously emerges in relation to discrimination on the grounds of specific individual characteristics and/or as a result of peer pressures (McPherson et al., 2001). SNS studies have shown that online social networks are so pervasive that they may well be considered as a crucial individual characteristic that prompt a negative bias towards non-users and vice versa[8].

On the other hand, *SN* players may get different payoffs when interacting with others of the same type depending on whether the interaction takes place online or offline. Several experiments, in fact, have shown that people behave online in a very peculiar way in respect to face-to-face interaction.

---

[8] For example, according to the *Social Recruiting Survey* conducted by Jobvite (2014), 92% of recruiters use social media for evaluating candidates. Furthermore, 94% use LinkedIn, 66% use Facebook and 52% use Twitter. Those who refer to Facebook, mostly use the platform to assess candidates' "cultural fit". People without Facebook pages, in particular, are viewed as "suspicious" by hiring managers, according to *Forbes* (Hill, 2012).

Kiesler et al. (1984) observed that computer-mediated communication entails anonymity, reduced self-regulation and reduced self-awareness. 'The overall weakening of self- or normative regulation might be similar to what happens when people become less self-aware and submerged in a group, that is, deindividuated' (p. 1126). Deindividuation has, in turn, been found to be conducive to disinhibition and lack of restraint (Diener, 1979). Siegel et al. (1983) found that people in computer-mediated groups were more aggressive than they were in face-to-face groups, as measured by uninhibited verbal behavior. Deregulation and disinhibition encourage "online incivility", which includes aggressive or disrespectful behaviors, vile comments, online harassment and hate speech.

Antoci et al. (2015b), Sabatini et al. (2015) and Sabatini and Sarracino (2015) argued that online incivility may be a major cause of frustration and dissatisfaction, which suggests that the benefits of the interaction between *SN* players could also be negative ($\beta < 0$) if the interaction takes place via SNS, and positive ($\gamma > 0$) if the interaction occurs face-to-face.

*3.2 Evolutionary Dynamics*

We assume that the adoption process of strategies *SN*, *NS* and *NP* is determined by the well-known replicator equations (see, e.g., Weibull 1995):

$$\begin{aligned} \dot{x}_1 &= x_1(EP_{SN} - \overline{EP}) \\ \dot{x}_2 &= x_2(EP_{NS} - \overline{EP}) \\ \dot{x}_3 &= x_3(EP_{NP} - \overline{EP}) \end{aligned} \qquad (1)$$

Where $\dot{x}_1$, $\dot{x}_2$, and $\dot{x}_3$ represent the time derivatives of the functions $x_1(t)$, $x_2(t)$, and $x_3(t)$, respectively, and:

$$\overline{EP} = x_1 EP_{SN} + x_2 EP_{NS} + x_3 EP_{NP}$$

is the population-wide average payoff of strategies.

Dynamics (1) are defined in the simplex *S* illustrated in Figure 1, where $x_1, x_2, x_3 \geq 0$ and $x_1 + x_2 + x_3 = 1$ hold. The vertices of *S*, that is the vectors $e_1 = (1,0,0)$, $e_2 = (0,1,0)$, and $e_3 = (0,0,1)$ correspond to the states in which all individuals adopt a unique strategy, respectively *SN*, *NS* or *NP*. We denote $e_i - e_j$ the edge of *S* joining $e_i$ with $e_j$; thus $e_1 - e_2$ is the edge where only strategies *SN* and *NS* are present in the population (see Figure 1), $e_1 - e_3$ is the edge where only strategies *SN* and *NP* are present, and $e_2 - e_3$ is the edge where only strategies *NS* and *NP* are present.

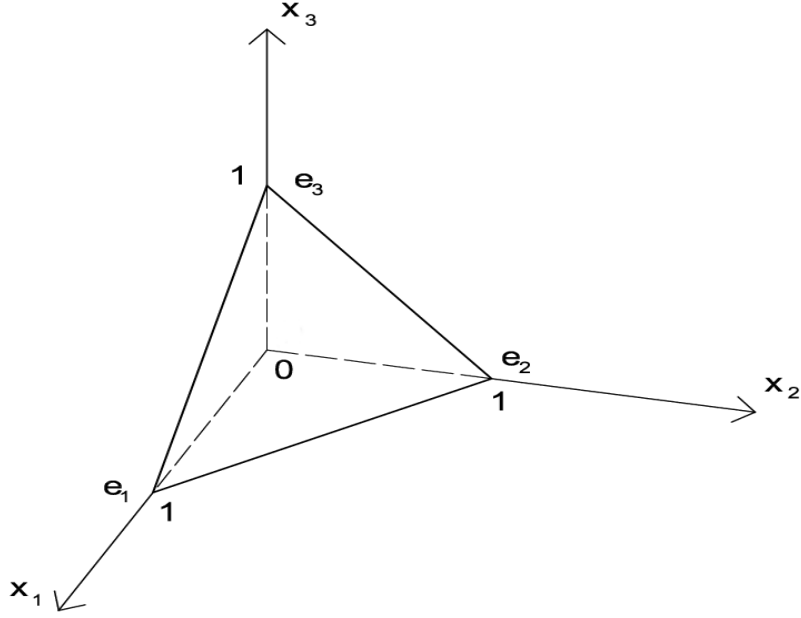

**Figure 1.** The simplex $S$, where $x_1, x_2, x_3 \geq 0$ and $x_1 + x_2 + x_3 = 1$ hold. The vertices $e_1 = (1,0,0)$, $e_2 = (0,1,0)$, and $e_3 = (0,0,1)$ correspond to the states in which all individuals adopt a unique strategy, respectively *SN*, *NS* or *NP*.

It is easy to check that dynamics (1) can be written in the following form (see, e.g., Bomze 1983):

$$\dot{x}_i = x_i(e_i \cdot A\boldsymbol{x} - \boldsymbol{x} \cdot A\boldsymbol{x}), \quad i = 1,2,3 \tag{2}$$

Where $\dot{x}_i$ represents the time derivative $dx_i/dt$ of $x_i(t)$, $\boldsymbol{x}$ is the vector $\boldsymbol{x} = (x_1, x_2, x_3)$, and $A$ is the payoff matrix:

$$A = \begin{pmatrix} (1-l)\alpha + \beta l^2 n^2 + \gamma l^2(1-n)^2 & (1-l)\alpha + \delta l^2(1-n) & (1-l)\alpha \\ (1-l)\alpha + \varepsilon l^2(1-n) & (1-l)\alpha + \eta l^2 & (1-l)\alpha \\ \alpha & \alpha & \alpha \end{pmatrix} \tag{3}$$

We will analyze dynamics (2) under the following assumptions:

***Assumption I***

$EP_{SN}(1,0) > EP_{NS}(1,0)$, that is $\beta n^2 + \gamma(1-n)^2 > \varepsilon(1-n)$: the *SN* strategy is better performing than the *NS* strategy in a social context where all individuals adopt the *SN* strategy (i.e., $x_1 = 1, x_2 = 0$).

*Assumption II*

$EP_{NS}(0,1) > EP_{SN}(0,1)$, that is $\eta > \delta(1-n)$: the *NS* strategy is better performing than the *SN* strategy in a social context where all individuals adopt the *NS* strategy (i.e., $x_1 = 0, x_2 = 1$).

*Assumption I* establishes a minimal condition for segregation. This condition is always satisfied if $\beta$ and $\gamma$, i.e., respectively the benefits that *SN* players get when they interact online and face-to-face with other *SN* players, are positive and $\varepsilon$, i.e., the reward that *NS* players get when interacting with *SN* players face-to-face, is negative or equal to zero. In this case, *SN* players will discriminate those who do not use online social networks, and *NS* players will not have any specific interest in engaging with them. More generally, *Assumption I* is satisfied if the value of $\varepsilon$ is lower enough than $\beta$ and $\gamma$.

*Assumption II* requires that the benefit $\delta$ obtained by *SN* players that meet *NS* individuals face-to-face is lower enough than the benefit obtained by *NS* players when they meet face-to-face individuals of their same type. This condition is certainly satisfied if $\eta \geq \delta$. In this case, *NS* players discriminate, in face-to-face encounters, those who adopt the *SN* strategy.

**4. Results**

It is well-known that dynamics (2) do not change if an arbitrary constant is added to each column of *A* (see, e.g., Hofbauer and Sigmund, 1988; p. 126). So, we can replace matrix *A*, in equations (2), with the following *normalized* matrix *B* with the first row made of zeros:

$$B = \begin{pmatrix} 0 & 0 & 0 \\ a & b & c \\ d & e & f \end{pmatrix} =$$

$$= \begin{pmatrix} 0 & 0 & 0 \\ \varepsilon l^2(1-n) - \beta l^2 n^2 - \gamma l^2(1-n)^2 & \eta l^2 - \delta l^2(1-n) & 0 \\ \alpha l - \beta l^2 n^2 - \gamma l^2(1-n)^2 & \alpha l - \delta l^2(1-n) & l\alpha \end{pmatrix} \quad (4)$$

According to *Assumptions I* and *II*, $a < 0$ and $b > 0$ hold. Furthermore, $f > 0$ always. The dynamic regimes that can be observed under *Assumptions I* and *II* can be classified taking into account the following results.

**Proposition 1**

1) *The stationary state $e_1$ –where all individuals play SN– is a sink (i.e., it is locally attractive) if the following condition holds (see matrix (4)):*

$$d = \alpha l - \beta l^2 n^2 - \gamma l^2 (1-n)^2 < 0 \quad i.e., \quad \alpha < \beta l n^2 + \gamma l (1-n)^2 \tag{5}$$

*While it is a saddle point if (5) does not hold (in such a case, the stable arm lies in the edge $e_1 - e_2$, while the unstable arm belongs to the edge $e_1 - e_3$).*

*2) The stationary state $e_2$ -where all individuals play NS- is a sink if the following condition holds (see matrix (4)):*

$$e - b = \alpha l - \eta l^2 < 0 \quad i.e., \quad \alpha < \eta l \tag{6}$$

*While it is a saddle point if (6) does not hold (in such a case, the stable arm lies in the edge $e_1 - e_2$, while the unstable arm belongs to the edge $e_2 - e_3$).*

*3) The stationary state $e_3$ –where all individuals play NP– is always a sink.*

**Proof**: See the mathematical appendix A.

Let us remember that:
- $1 - l$ and $l$ are, respectively, the probabilities to be a *Wmode*-individual or an *Lmode*-individual (playing either the *SN* or the *NS* strategy), while $n$ is the probability for an individual playing the *SN* strategy to be interacting online via SNS.
- The parameter α measures the payoff of a *Wmode*-individual.
- The parameter β measures the benefits for an *Lmode*-individual choosing the *SN* strategy when she is interacting via a SNS with individuals of the same type.
- The parameters γ and δ measure the benefits of an *Lmode*-individual choosing the *SN* strategy when she is interacting via face-to-face encounters with, respectively, individuals of the same type and with *Lmode*-individuals choosing the *NS* strategy.
- The parameters η and ε measure the benefits of an *Lmode*-individual adopting the *NS* strategy when she is interacting face-to-face with, respectively, individuals of the same type and with *Lmode*-individuals adopting *SN*.

Note that conditions (5) and (6) are simultaneously satisfied if the value of the parameter $\alpha$ – measuring the (constant) payoff of the *NP* strategy is low enough. Differently from $e_1$ and $e_2$, the stationary state $e_3$ is always a sink, whatever the value of the parameter $\alpha > 0$ is.

When the pure population states $e_1$, $e_2$, and $e_3$ are sinks, they are also Nash equilibria (see, e.g., Weibull, 1995). In such a case, they can be interpreted as stable social conventions representing self-enforcing configurations of the social environment.

In our model, individuals' welfare evaluated at $e_1$, $e_2$, and $e_3$ is measured, respectively, by:

$$EP_{SN}(1,0) = (1-l)\alpha + \beta l^2 n^2 + \gamma l^2 (1-n)^2$$
$$EP_{NS}(0,1) = (1-l)\alpha + \eta l^2$$
$$EP_{NP} = \alpha$$

The following proposition deals with Pareto dominance relationships among the stationary states $e_1$, $e_2$, and $e_3$.

**Proposition 2**. *The stationary state $e_1$ –where all individuals play SN– Pareto dominates the stationary state $e_2$ –where all individuals play NS- (i.e., $EP_{SN}(1,0) > EP_{NS}(0,1)$) if:*

$$\eta < \beta n^2 + \gamma(1-n)^2 \qquad (7)$$

*and Pareto dominates the stationary state $e_3$ –where all individuals play NP- (i.e., $EP_{SN}(1,0) > EP_{NP}$) if:*

$$\alpha < \beta l n^2 + \gamma l (1-n)^2 \qquad (8)$$

*The stationary state $e_2$ –where all individuals play NS– Pareto dominates the stationary state $e_3$ – where all individuals play NP- (i.e., $EP_{NS}(1,0) > EP_{NP}$) if:*

$$\alpha < \eta l \qquad (9)$$

**Proof**: Straightforward.

It is important to note that (8) and (9) coincide, respectively, with the stability conditions (5) and (6). Therefore, if $e_1$ and $e_2$ are sinks, then they Pareto dominate the stationary state $e_3$ –where individuals withdraw from social participation. This implies that, in the context in which at least one of the stationary states $e_1$ and $e_2$ are sinks, the stationary state $e_3$ (which is always locally attractive) can be interpreted as a *social poverty trap*. In the trap, everyone withdraws from social participation and devotes all of their available time to work and "private" activities, e.g., work and consumption that do not entail any significant social relationship. The "social poverty" that derives from this situation –manifesting, for example, in the scarcity of participation opportunities and in the strengthening of materialistic values– makes social interaction difficult and unrewarding.

Also note that the Pareto dominance relationship between $e_1$ and $e_2$ (see (7) does not depend on the stability conditions (5) and (6), and, consequently, $e_1$ may Pareto dominate $e_2$ or vice versa, independently from their stability properties.

The following proposition concerns the existence and stability properties of the other possible stationary states of dynamics (2), that is, the stationary states where at least two, among the available strategies, are adopted by (strictly) positive shares of the population.

**Proposition 3**

*1) A unique stationary state in the interior of S (i.e., with $x_i > 0$ all $i = 1,2,3$), where all strategies are played, exists if:*

$$ae - bd = l^3\{\varepsilon(1-n)[\alpha - \delta l(1-n)] + \alpha\delta(1-n)\} + $$
$$+l^3\{(\eta l - \alpha)[\beta n^2 + \gamma(1-n)^2] - \alpha\eta\} > 0 \qquad (10)$$

*Such a stationary state is always a source (i.e., it is repulsive). If condition (10) does not hold, then no stationary state exists in the interior of S.*

*2) A unique stationary state always exists in the edge $e_1 - e_2$ (not coinciding with either $e_1$ or $e_2$) of the simplex S (see Figure 1); it is a saddle point (with unstable manifold lying in $e_1 - e_2$) if the stationary state in the interior of S exists (see point one of this proposition), otherwise it is a source.*

*3) A unique stationary state exists in the edge $e_1 - e_3$ if $d < 0$ (see condition (5), and it is always a saddle point (with unstable manifold lying in $e_1 - e_3$). If $d \geq 0$, then no stationary state exists in $e_1 - e_3$.*

*4) A unique stationary state exists in the edge $e_2 - e_3$ if $e - b < 0$ (see condition (6), and it is always a saddle point (with unstable manifold lying in $e_2 - e_3$). If $e - b \geq 0$, then no stationary state exists in $e_2 - e_3$.*

**Proof**: See the mathematical appendix A.

# 5. Classification of Dynamic Regimes

Bomze (1983) provided a complete classification of two-dimensional replicator equations. The above propositions allow us to select, among all of the phase portraits illustrated in Bomze's paper, those that can be observed under the dynamics (2). In Figures 2-8, sinks (i.e., attractive stationary states) are indicated by full dots, sources (i.e., repulsive stationary states) are indicated by open dots and saddle points by drawing their stable and unstable branches. The basins of attraction of $e_1$, $e_2$, and $e_3$ are, respectively, in yellow, blue and pink. According to Proposition 3 (and to Bomze's classification), every trajectory starting from an initial distribution of strategies $x_1(0), x_2(0)$, and $x_3(0)$ –neither belonging to a one-dimensional stable manifold of a saddle point nor coinciding with a stationary state where more than one strategy is adopted– approaches one of the pure population stationary states $e_1$, $e_2$, and $e_3$. In the following subsections, we will present the complete classification of the possible dynamics regimes that can be observed under equations (2).

## 5.1. Regime One: Conditions (5) and (6) Hold

In this context, all of the vertices $e_1 = (1,0,0)$, $e_2 = (0,1,0)$, and $e_3 = (0,0,1)$ are simultaneously attractive and the regimes illustrated in Figures 2 and 3 can be observed. The former – corresponding to the phase portrait number 35 (PP#35) in Bomze's (1983) classification– occurs when $ae - bd \leq 0$ (i.e., when a stationary state in the interior of $S$ does not exist, see condition (10), while the latter –corresponding to PP#7– occurs when $ae - bd > 0$.

In this context, the stationary state $e_3 = (0,0,1)$ –where all the individuals play the *NP* strategy– is Pareto dominated by the other locally attracting stationary states. This suggests that the *NP* strategy can be interpreted as an adaptive behavior that agents may want to play to protect themselves from situations of relational poverty and decay of the surrounding social environment. As clearly illustrated in Figures 2 and 3, these regimes are strongly path dependent. If the initial distribution of the different forms of participation is close enough to $e_1 = (1,0,0)$, i.e., if $x_1(0)$ is high enough and $x_2(0)$ and $x_3(0)$ are low enough, then the economy converges to $e_1$, where all individuals adopt the *SN* strategy. On the other hand, if the initial distribution is close enough to $e_2$ or $e_3$, then the economy converges to $e_2$ or $e_3$.

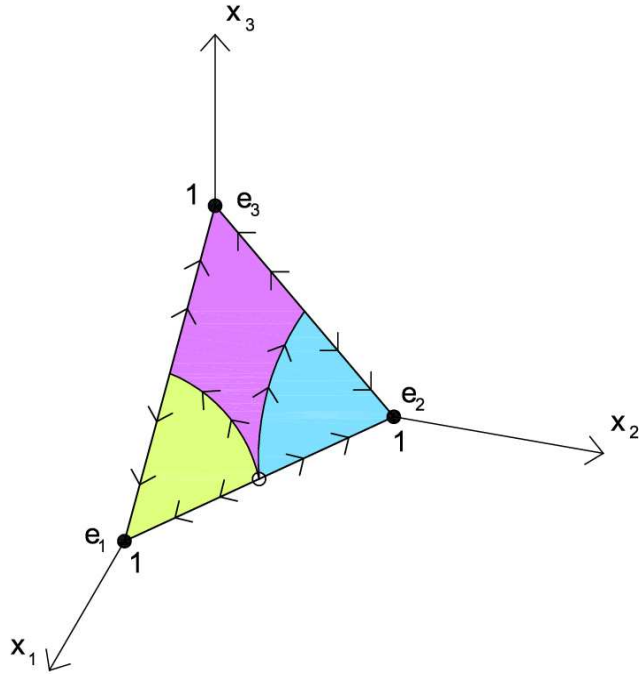

**Figure 2.** Dynamic regime in which all of the vertices $e_1 = (1,0,0)$, $e_2 = (0,1,0)$, and $e_3 = (0,0,1)$ are simultaneously attractive and a stationary state in the interior of $S$ does not exist. The basins of attraction of $e_1$, $e_2$ and $e_3$ are, respectively, in yellow, blue and pink.

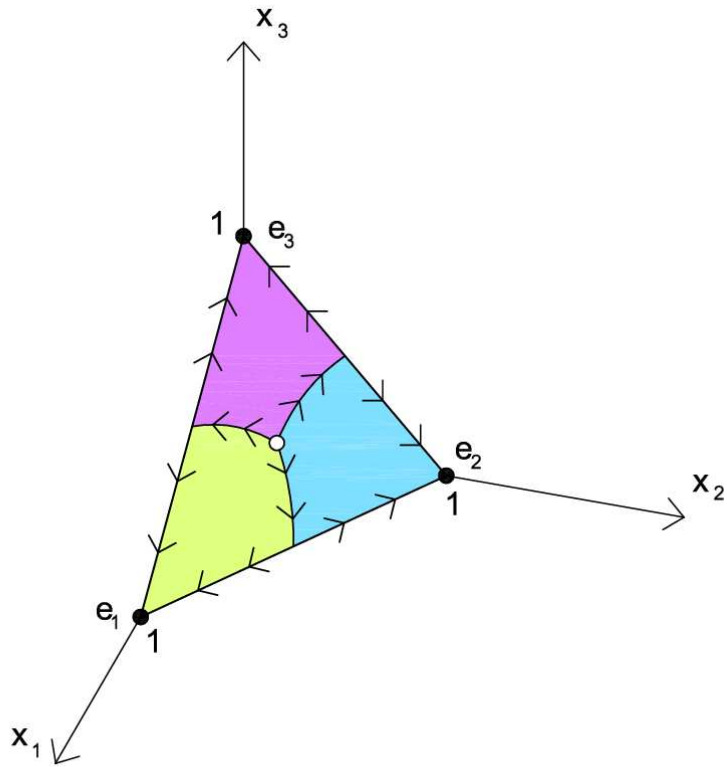

**Figure 3.** Dynamic regime in which all of the vertices $e_1 = (1,0,0)$, $e_2 = (0,1,0)$, and $e_3 = (0,0,1)$ are simultaneously attractive and a stationary state in the interior of $S$ exists. The basins of attraction of $e_1$, $e_2$ and $e_3$ are, respectively, in yellow, blue and pink.

## 5.2. Regime Two: Condition (5) Holds, But (6) Does Not Hold

In this context, the vertices $e_1 = (1,0,0)$ and $e_3 = (0,0,1)$ are attractive, while $e_2 = (0,1,0)$ is a saddle point. The regimes are illustrated in Figures 4 and 5.

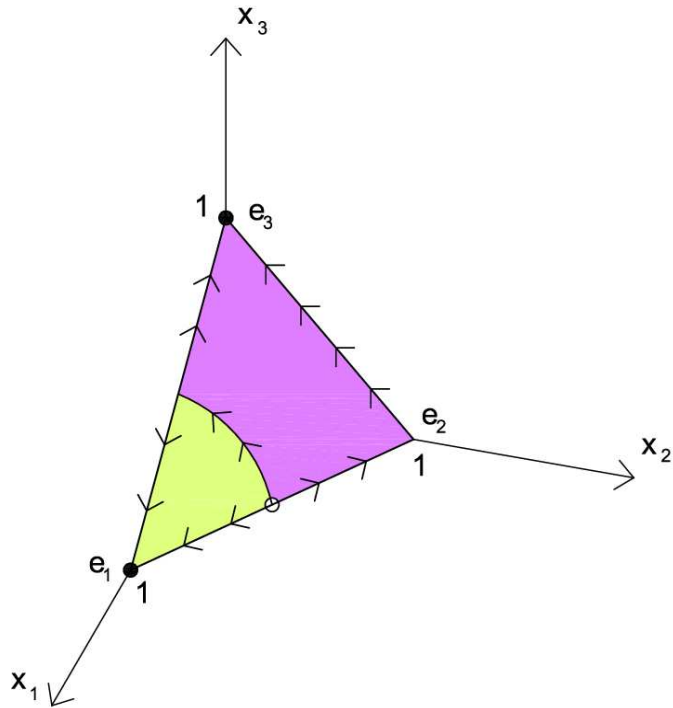

**Figure 4**. Dynamic regime in which only the vertices $e_1 = (1,0,0)$ and $e_3 = (0,0,1)$ are attractive, and a stationary state in the interior of $S$ does not exist. The basins of attraction of $e_1$ and $e_3$ are, respectively, in yellow and pink.

The regime in Figure 4 –corresponding to PP#37 of Bomze's classification– occurs when $ae - bd \leq 0$ (i.e., when a stationary state in the interior of $S$ does not exist, see condition (10), while the latter –corresponding to PP#9– occurs when $ae - bd > 0$. In this context, the stationary state $e_3 = (0,0,1)$ –where all the individuals play the *NP* strategy– is Pareto dominated by $e_1 = (1,0,0)$ – where all the individuals play the *SN* strategy. Furthermore, the stationary state $e_2 = (0,1,0)$ – where all the individuals play the *NS* strategy– is Pareto dominated by both the stationary states $e_1 = (1,0,0)$ and $e_3 = (0,0,1)$.

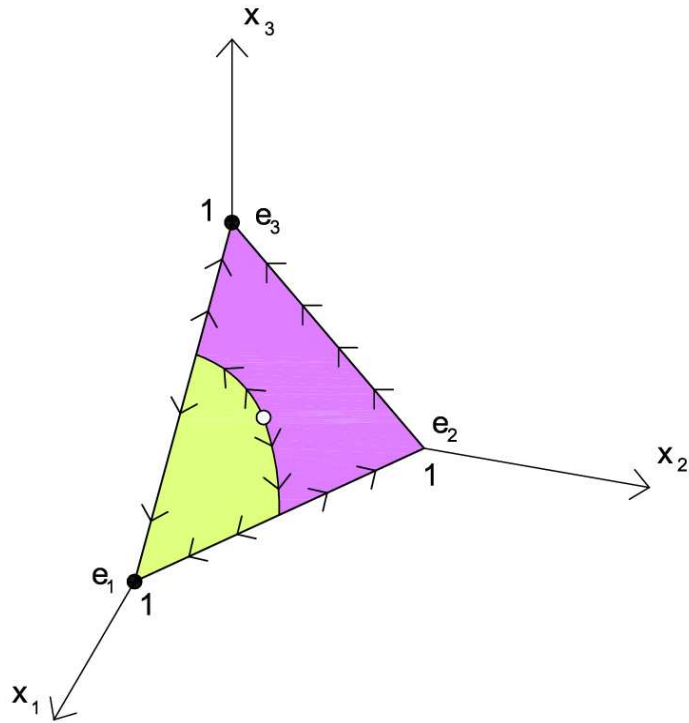

**Figure 5**. Dynamic regime in which only the vertices $e_1 = (1,0,0)$ and $e_3 = (0,0,1)$ are attractive, and a stationary state in the interior of $S$ exists. The basins of attraction of $e_1$ and $e_3$ are, respectively, in yellow and pink.

### 5.3. Regime Three: Condition (6) Holds, But (5) Does Not Hold

In this context, the vertices $e_2 = (0,1,0)$ and $e_3 = (0,0,1)$ are attractive, while $e_1 = (1,0,0)$ is a saddle point. The regimes are illustrated in Figures 6 and 7. The regime in Figure 6 –corresponding to PP#37 of Bomze's classification– occurs when $ae - bd \leq 0$ (i.e., when a stationary state in the interior of $S$ does not exists, see condition (10), while the latter –corresponding to PP#9– occurs when $ae - bd > 0$. In this context, the stationary state $e_3 = (0,0,1)$ –where all the individuals play the *NP* strategy– is Pareto dominated by $e_2 = (0,1,0)$ –where all the individuals play the *NS* strategy. Furthermore, the stationary state $e_1 = (1,0,0)$ –where all the individuals play the *SN* strategy– is Pareto dominated by both the stationary states $e_2 = (0,1,0)$ and $e_3 = (0,0,1)$.

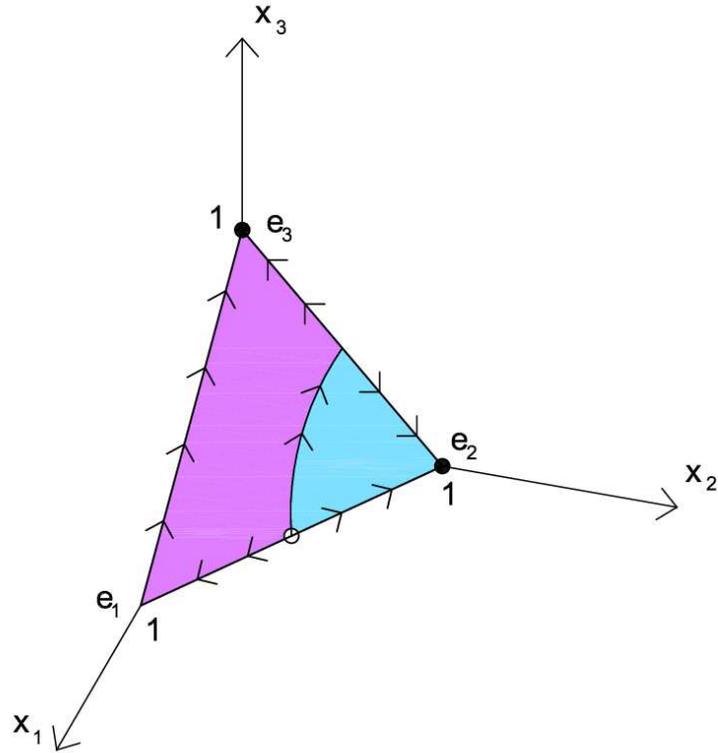

**Figure 6**. Dynamic regime in which only the vertices $e_2 = (0,1,0)$ and $e_3 = (0,0,1)$ are attractive, and a stationary state in the interior of $S$ does not exist. The basins of attraction of $e_2$ and $e_3$ are, respectively, in blue and pink.

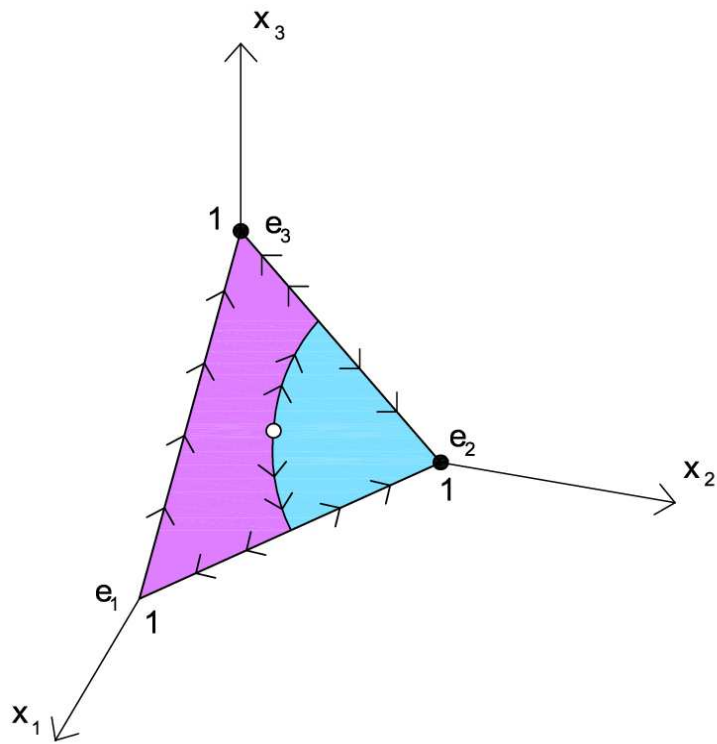

**Figure 7**. Dynamic regime in which only the vertices $e_2 = (0,1,0)$ and $e_3 = (0,0,1)$ are attractive, and a stationary state in the interior of $S$ exists. The basins of attraction of $e_2$ and $e_3$ are, respectively, in blue and pink.

## 5.4. Regime Four: Neither Condition (5) Nor (6) Hold

In this context, $ae - bd \leq 0$ always holds (i.e., a stationary state in the interior of $S$ does not exist), and the unique dynamic regime that can be observed is illustrated in Figure 8 –corresponding to PP#42 of Bomze's classification. In this regime, the unique attractive stationary state is $e_3 = (0,0,1)$, where all individuals withdraw from social participation, which Pareto dominates both the stationary states $e_1 = (1,0,0)$ and $e_2 = (0,1,0)$.

This extreme scenario may be interpreted as the result of exogenous conditions of social decay, which make social participation (in any form) poorly rewarding. For instance, the scarcity of infrastructures for face-to-face interactions (e.g., meeting places such as public parks, theatres, clubs, associations) lowers the reward provided by the *NS* strategy. Furthermore, the poverty of technological infrastructures for fast Internet access lowers the reward associated with the *SN* strategy.

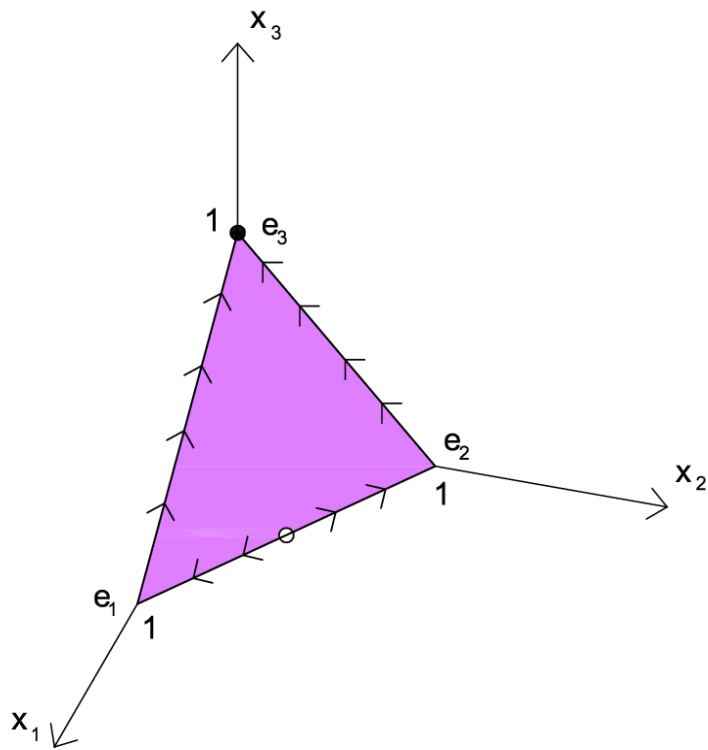

**Figure 8**. Dynamic regime in which only the vertex $e_3 = (0,0,1)$ is attractive. Its basin of attraction is in pink.

## 5.5 Discrimination and the Social Poverty Trap

The classification of dynamic regimes illustrated in Figures 2-8 suggests that the structure of the basin of attraction of the social poverty trap $e_3$ crucially depends on the propensity for discrimination of the two sub-populations of socially active individuals. The higher the propensity for discrimination, the greater the probability that individuals will ultimately segregate themselves, making society fall into the trap.

In fact, the less gratifying the interaction between *SN* and *NS* players, the more attractive the social isolation strategy *NP* becomes. If the rewards $\delta$ and $\varepsilon$ that *SN* and *NS* players get when they interact face-to-face is particularly low, then they both might be tempted to withdraw from social interaction, whatever the initial share of the sub-population adopting either *SN* or *NS* is. Notice that condition (10) (when it holds, then a stationary state in the interior of the simplex *S* exists) is never satisfied if the rewards $\delta$ and $\varepsilon$ are negative and low enough. In such a context, if the stationary states $e_1$ and $e_2$ are attractive (see the regime shown in Figure 2), then the basin of attraction of the social poverty trap $e_3$ is so large that it includes the area bordering the edge $e_1 - e_2$, where the *NP* strategy is almost extinct, the majority of the population socially participates and the two strategies of social participation (*NS* and *SN*) are uniformly distributed.

However, the basin of attraction of the social poverty trap $e_3$ does not include the areas in close proximity to the edge $e_1 - e_2$ if the rewards $\delta$ and $\varepsilon$ are high enough and, therefore, condition (10) is satisfied (see Figure 3). This result suggests that, if the two sub-populations of *SN* and *NS* players have a limited tendency to discriminate each other –which happens if the rewards $\delta$ and $\varepsilon$ that the two types of player get when they interact face-to-face are high enough– then society will less likely fall into the social poverty trap in the cases in which the initial level of social participation is high, even if the two strategies *NS* and *SN* are uniformly distributed, as it happens in the dynamic regime illustrated in Figure 3. On the other hand, when the reward given by the interaction between *SN* and *NS* players is particularly low, the two strategies ultimately may crowd out each other. A similar crowding out effect also applies to the dynamic regimes illustrated in Figures 4, 5, 6 and 7. In these cases, the basin of attraction of the social poverty trap $e_3$ is so large that it also includes the areas in close proximity to the edge $e_1 - e_2$. This means that society can converge to $e_3$ even if the initial share of the sub-population adopting the social participation strategies *SN* and *NS* is particularly high.

## 6. Supplementary Result: A Prediction of the Model

There is growing empirical evidence showing that face-to-face interaction is associated with higher levels of well-being than SNS-mediated interactions. Using Italian cross-sectional data, Sabatini and Sarracino (2014b) found that subjective well-being is positively correlated with face-to-face encounters and negatively correlated with SNS-mediated interactions. Helliwell and Huang (2013) reached a similar conclusion by comparing the well-being effects of online and offline friendships in a Canadian sample. Kross et al. (2013) examined this issue using experience sampling. The authors text-messaged people five times per day for two weeks to test how offline and Facebook-mediated interactions correlate with aspects of subjective well-being (SWB). Results indicate that Facebook use predicts negative shifts in SWB, while face-to-face interactions show no significant effect. Based on a survey conducted on a representative sample of 2,000 French Facebook users, Pénard and Mayol (2015) found that Facebook interferes with subjective well-being through its effects on friendships and self-esteem. Their results show that people who also use the network to seek social approval in the form of more *Likes* tend to be more unsatisfied with their life. Similarly, Sabatini and Sarracino (2015b) drew on Italian representative data to show that the use of SNS is associated with lower levels of satisfaction with respondents' income, differently from face-to-face interactions, thereby suggesting that the use of online networks can raise material aspirations with detrimental effects for SWB.

Overall, the empirical evidence suggests the convenience of further analyzing the dynamics occurring in the region of the simplex where:

$$EP_{NS}(x_1, x_2) > EP_{SN}(x_1, x_2)$$

In this region, the reward provided by a strategy of social participation exclusively based on face-to-face interactions is higher than the benefits associated with the use of SNS (the *SN* strategy). The following proposition allows the prediction of the possible evolution of the shares of the population $x_1, x_2, x_3$ adopting the three strategies in a society, starting from an initial configuration of payoffs that are consistent with the evidence mentioned above, i.e., where:

$$EP_{NS}(x_1(0), x_2(0)) > EP_{SN}(x_1(0), x_2(0))$$

**Proposition 4**

*The set in which*

$$EP_{NS}(x_1, x_2) > EP_{SN}(x_1, x_2)$$

*holds (where the payoff of strategy SN is lower than that of strategy NS) and the set in which*

$$EP_{NS}(x_1, x_2) < EP_{SN}(x_1, x_2)$$

*holds (where the payoff of strategy SN is higher than that of strategy NS) are invariant under dynamics (2). That is, every trajectory starting from the former cannot enter the latter, and vice versa.*

**Proof**: See the mathematical appendix B.

Proposition 4 states that, if the payoff associated with *NS* is initially higher than the one associated with *SN*, then it will always be higher than that provided by *SN*, unless exogenous perturbations will change the model's parameters. As a result, the economy cannot converge to the stationary state $e_1 = (1,0,0)$, where all individuals adopt the *SN* strategy if starting from the region in which $EP_{NS} > EP_{SN}$ holds. This means that almost all of the trajectories starting from such a region will converge to $e_2$ –where all individuals socially participate by exclusive means of face-to-face interactions- or to $e_3$ –where nobody participates. Only one trajectory can reach the edge $e_2 - e_3$ where the *NS* and the *NP* strategies coexist. In any case, the analysis of dynamics suggests that society will converge to equilibria where no one adopts the *SN* strategy.

## 7. Conclusions

In this paper, we developed an evolutionary game model to study the dynamics of different modes of interaction in an era that is characterized by a steep rise in the use of online social networks and the supposed decline in face-to-face social participation. In our framework, individuals can choose to withdraw from social relations or interact with others by means of SNS and/or face-to-face encounters. The analysis showed that, depending on the configuration of payoff and the initial distribution of the various modes of participation in the population, different Nash equilibria could be reached. If we allow a configuration of payoff that is compatible with individuals' preference for similar others, then discrimination will lead to the segregation of the three sub-populations accounted for in the analysis and, ultimately, to the survival of only one of the three. Every trajectory starting from an initial distribution of strategies neither belonging to a one-dimensional stable manifold of a saddle point nor coinciding with a stationary state, where more than one strategy is adopted, will approach one of the pure population stationary states.

If the reward for social withdrawal is low enough, then the stationary states, where all individuals play one of the two strategies of participation, *e₁* and *e₂*, are locally attractive. In this case, they both Pareto dominate the stationary state, where everyone withdraws from social interaction, *e₃*. However, there is no Pareto dominance relationship between *e₁* and *e₂*.

If $e_1$ and $e_2$ are attractive, then the former can Pareto dominate the latter or vice versa, but both the equilibria Pareto dominate the social poverty trap $e_3$. The dynamic regimes are strongly path dependent. If the initial distribution of the three strategies is close enough to $e_1$, then the economy will converge to $e_1$. The same can be said for $e_2$ and $e_3$. The social poverty trap $e_3$, on the other hand, is always a sink, whatever the payoff of social withdrawal is. In this scenario, the withdrawal from social participation can be interpreted as a defensive behavior in the sense theorized by Hirsch (1976). Individuals, in fact, might want to cope with the deterioration in the social environment surrounding them and/or the increasing busyness related to their material aspirations by choosing to limit their social relationships to a minimum. This result is related to previous research that studied how growth may cause negative externalities on social relationships and social cohesion (Putnam, 2000; Bartolini and Bonatti, 2008; Antoci et al., 2007, 2012b, 2013; Bartolini et al., 2013; Bartolini and Sarracino, 2015a). These studies claimed that the rise in material aspirations and the need to work more might tighten time constraints, causing deterioration in the social environment and prompting a gradual withdrawal from face-to-face interactions

Social withdrawal is self-feeding, in that the higher the share of the population renouncing to social participation, the poorer the social environment becomes, for example, in terms of social engagement opportunities. People playing the *NP* strategy will ultimately decide to segregate themselves from the rest of the population.

In all of the possible cases, corresponding to the stationary states $e_1$, $e_2$ and $e_3$, the segregation entailed by individuals' tendency for discrimination will lead to the survival of only one of the initial sub-populations.

The model also allowed us to study the future of social participation in a world in which social interaction via online networks is less rewarding than offline interaction. This scenario is particularly interesting as it is consistent with findings from the most recent empirical studies comparing the effect of face-to-face and SNS-mediated interactions on individuals' well-being. Our results suggest that dynamics starting from this scenario will lead the *SN* strategy to extinction, which entails that Facebook and alike will disappear.

## Mathematical Appendix A

Dynamics (2) is equivalent (see Hofbauer, 1981) to the Lotka-Volterra system:

$$\dot{X} = X(a + bX) \tag{11}$$
$$\dot{Y} = Y(d + eX + fY) \tag{12}$$

via the coordinate change:

$$x_1 = \frac{1}{1+X+Y}, \quad x_2 = \frac{X}{1+X+Y}, \quad x_3 = \frac{Y}{1+X+Y} \tag{13}$$

From which $X = x_2/x_1$ and $Y = x_3/x_1$.

Please note that, by the coordinate change (13), the edge $e_1 - e_2$ of the simplex $S$ (see Figure 1) corresponds to the positive semi-axis $Y = 0$ of the plane $(X,Y)$, the edge $e_1 - e_3$ corresponds to the positive semi-axis $X = 0$ and the vertex $e_1$ corresponds to the point $(X,Y)=(0,0)$ (see Figure 9).

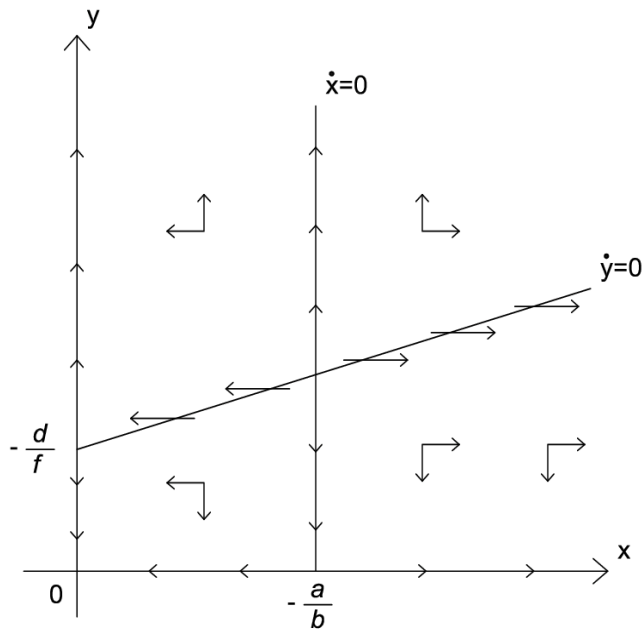

**Figure 9.** Arrow diagram of the Lotka-Volterra system. The edge $e_1 - e_2$ of the simplex $S$ corresponds to the positive semi-axis $Y = 0$ of the plane $(X,Y)$, the edge $e_1 - e_3$ corresponds to the positive semi-axis $X = 0$, and the vertex $e_1$ corresponds to the point $(X,Y)=(0,0)$. The set in which $EP_{SN} > EP_{NS}$ holds coincides with the region on the left of the vertical straight line $X = -a/b$.

According to equation (11), $\dot{X} = 0$ holds along the axis $X = 0$ and along the vertical straight line $X = -a/b > 0$; furthermore $\dot{X} > 0$ ($\dot{X} < 0$) holds on the right (respectively, on the left) of $X = -a/b$. According to equation (12), $\dot{Y} = 0$ holds along the axis $Y = 0$ and along the straight line $Y = -d/f - (e/f)X$. Furthermore, $\dot{Y} > 0$ ($\dot{Y} < 0$) holds above (respectively, below) $Y = -d/f - (e/f)X$.

Remembering that $a < 0, b > 0,$ and $f > 0$, we have that a unique stationary state with $X > 0$ and $Y > 0$, $(\bar{X}, \bar{Y}) = \left(-a/b, -d/f + (ae)/(bf)\right)$, exists if and only if $ae > bd$ (condition (10) of Proposition 3). The Jacobian matrix of system (11)-(12), evaluated at $(\bar{X}, \bar{Y})$, is a triangular matrix:

$$J(\bar{X}, \bar{Y}) = \begin{pmatrix} b\bar{X} & 0 \\ \bar{Y} & f\bar{Y} \end{pmatrix}$$

With eigenvalues $b\bar{X} > 0$ (in direction of $= -a/b$) and $f\bar{Y} > 0$. So $(\bar{X}, \bar{Y})$ is always a repulsive node (this completes the proof of point one of Proposition 3).

By following similar steps, it is easy to check that:

1) The Lotka-Volterra system (11)-(12) always admits a unique stationary state $(X, Y) = (-a/b, 0)$, with $-a/b > 0$, belonging to the positive semi-axis $Y = 0$ (corresponding to the edge $e_1 - e_2$ of the simplex $S$, see Figure 1). Such a stationary state is a saddle point (with unstable manifold lying in $Y = 0$, and stable manifold lying in $X = -a/b$) if the internal stationary state $(\bar{X}, \bar{Y})$ exists; otherwise it is a source (point two of Proposition 3).

2) The Lotka-Volterra system (11)-(12) admits a unique stationary state $(X, Y) = (0, -d/f)$, with $-d/f > 0$, belonging to the positive semi-axis $X = 0$ (corresponding to the edge $e_1 - e_3$ of the simplex $S$) if $d < 0$. Such a stationary state is always a saddle point with unstable manifold lying in $X = 0$. If $d \geq 0$, then no stationary state with $Y > 0$ exists in the positive semi-axis $X = 0$ (point three of Proposition 3).

3) The state $(X, Y) = (0,0)$ (corresponding to the vertex $e_1$ of the simplex $S$, see Figure 1) is always a stationary state; it is a saddle point (with unstable manifold lying in $X = 0$, and stable manifold lying in $Y = 0$) if $d \geq 0$ (i.e., if the stationary state in the semi-axis $X = 0$ does not exist, see point 2 above), otherwise it is a sink (point one of Proposition 1).

The stability properties of the stationary states $e_2$ and $e_3$ (points two to three of Proposition 1), and the existence and stability properties of the stationary state belonging to the edge $e_2 - e_3$ (point four of Proposition 3)[9] can be easily analyzed by applying Propositions 1, 2 and 5 in Bomze (1983), who provided a complete classification of two-dimensional replicator equations.

**Mathematical Appendix B**

The condition:

$$EP_{SN}(x_1, x_2) > EP_{NS}(x_1, x_2)$$

can be written as follows:

$$ax_1 + bx_2 < 0,$$
$$bx_2 < -ax_1,$$
$$X < -\frac{a}{b},$$

where $X = x_2/x_1$. Consequently, in the positive quadrant of the plane $(X, Y)$, the set in which $EP_{SN} > EP_{NS}$ holds coincides with the region on the left of the vertical straight line (see Figure 9):

$$X = -\frac{a}{b} > 0 \qquad (14)$$

Along the straight line (14), $\dot{X} = 0$ holds, while the set in which $EP_{SN} < EP_{NS}$ holds corresponds to the region on the right of (14). Since (14) cannot be crossed by trajectories (see Figure 9), the two regions separated by (14) are invariant. Consequently, every trajectory starting from the region in which $EP_{SN} < EP_{NS}$ cannot converge to the stationary state $(X, Y) = (0,0)$, which corresponds to the stationary state $e_1 = (1,0,0)$. This completes the proof of Proposition 4.

---

[9] Such stationary states do not correspond to stationary states of the Lotka-Volterra system (11)-(12).